\documentclass{elsart1p}
\def\elsartstyle{%
    \def\normalsize{\@setfontsize\normalsize\@xiipt{14.5}}
    \def\small{\@setfontsize\small\@xipt{13.6}}
    \let\footnotesize=\small
    \def\large{\@setfontsize\large\@xivpt{18}}
    \def\Large{\@setfontsize\Large\@xviipt{22}}
    \skip\@mpfootins = 18\p@ \@plus 2\p@
    \normalsize
}
\def\file#1{\texttt{#1}}
\usepackage{graphicx}
\usepackage[figuresright]{}

\begin{document}

\begin{frontmatter}
\title{The use of neural networks to probe the structure of the nearby universe}

\author{Raffaele d'Abrusco }
\address{Dept. of Physical Sciences, University Federico II in Napoli, via Cinthia 6, 80126-Napoli, Italy
\& Institute of Astronomy, Cambridge, UK}
\author{Giuseppe Longo \& Maurizio Paolillo}
\address{Dept. of Physical Sciences, University Federico II in Napoli, via Cinthia 6, 80126-Napoli, Italy
\& INFN, Napoli Unit, via Cinthia 6, 80126-Napoli, Italy}
\author{Massimo Brescia, Elisabetta De Filippis \& Antonino Staiano}
\address{INAF, Napoli Unit, via Moiariello 16, 80131-Napoli, Italy}
\author{Roberto Tagliaferri}
\address{Dept. of Mathematics and Informatics, University of Salerno, Fisciano, Italy}

\ead{dabrusco@na.infn.it}

\begin{abstract}
In the framework of the European VO-Tech project, we are implementing new
machine learning methods specifically tailored to match the needs of 
astronomical data mining. 
In this paper, we shortly present the methods and discuss an application to the
Sloan Digital Sky Survey public data set. In particular, we discuss some 
preliminary results on the 3-D taxonomy of the nearby ($z<0.5$) universe. 
Using neural networks trained on the available spectroscopic base of 
knowledge we derived distance estimates for $\sim 30$ million galaxies
distributed over $\sim 8,000$ sq. deg. 
We also use unsupervised clustering tools to investigate whether it is possible
to characterize in broad morphological bins the nature of each object
and produce a reliable list of candidate AGNs and QSOs.
\end{abstract}

\end{frontmatter}

\section{Introduction}\label{intro}
Modern photometric digital surveys performed with dedicated instruments 
are in fact producing high quality multiband data at a huge rate, and a 
new generation of software tools is being developed to merge databases and to
extract patterns, trends, etc. 
Furthermore, due the ongoing efforts 
to build a Virtual Observatory (VO) infrastructure (cf. the European VO-Tech, 
\cite{votech}), everyone will have at his 
fingertips the possibility to extract and manipulate such multi-wavelength, 
multi-epoch, multi-instrument datasets of huge dimensionality. 
The extraction of useful information from such datasets 
imposes to abandon old conceptual schemes largely based on the 3-D
visualization capability of human minds,  and to introduce in 
the astronomical practice fine tuned, machine learning methods for 
statistical pattern recognition, classification and visualization 
\cite{longo_2005}. In this context, however, astronomical data pose non 
trivial problems due to the fact that they are usually highly degenerate, 
present strong non linear correlations among parameters
and are plagued by large fraction of missing data and upper limits.

In what follows we shall shortly outline the first results of our 
ongoing effort to implement and apply a new generation of tools to the 
construction of a 3-D taxonomy of the nearby universe with a 
characterization of galaxy types in a few, broadly defined, catagories: 
normal (early and late type) galaxies, AGN, QSO, etc.

For all the experiments described below, we used the Sloan Digital Sky Survey 
Data Release 4 and/or 5 (hereafter SDSS4/5; \cite{SDSS5}) which provides 
photometric data in 5 bands for several hundred million galaxies distributed 
over $\sim 8,000$ square degrees, together with additional spectroscopic 
information for a subsample (hereafter SpS) of $\sim 10^6$ galaxies. 
Such spectroscopic information include, among other things, redshifts
and a spectral classification index derived from an eigenvector analysis 
which allows to disentangle among normal galaxies (SP2),
stars (SP1), nearby AGN (SP3), distant AGN (SP4), late type stars (SP6).
This spectroscopic base of knowledge can be used as a training ground
for interpolative and supervised methods as well as a validation ground for
unsupervised ones.  
\section{The photometric redshift}\label{redshifts}
One of the main sources of degeneracy in the photometric properties of
galaxies is the lack of information on their distances and therefore any
attempt to partition the parameter space requires first the derivation 
of distance estimates. 
In order to evaluate photometric redshifts we made use of an improved version 
\cite{dabrusco_2006} of the Neural Networks (NNs) method presented in 
\cite{tagliaferri_2003}. 
Both steps are accomplished using NNs architecture known as MLP 
(Multi Layer Perceptron).

We adopted a two steps approach based on the use of MLP \cite{bishop_1995,duda}.
First we trained a MLP to recognize nearby (id est with redshift $z<0.25$) 
and distant ($z>0.25$) objects, then we trained two separate MLPs
to work in the two different redshift regimes. 
Such approach finds support in the fact that in the SDSS-5 catalogue, 
the distribution of galaxies inside the two different redshift intervals 
is dominated by two different galaxy populations: the Main Galaxy (MG) sample 
in the nearby region, and the Luminous Red Galaxies (LRG) in the distant one. 
The use of two separate networks ensures that the NNs achieve a good generalization
capabilities in the nearby sample, leaving the biases mainly in the distant one.

To perform the separation between MG and LRG objects, we extracted from the 
SDSS-4 SpS training, validation and test sets weighting, respectively, $60\%$, 
$20\%$ and $20\%$ of the total number of objects (449,370 galaxies).
The resulting test set, therefore, consisted of 89,874  randomly extracted 
objects. 

After the first network has classified the objects in nearby and distant ones, 
the derivation of the photometric redshifts is performed separately in the two 
regimes.
Since NNs are excellent at interpolating data but very poor in extrapolating them,
in order to minimize the systematic errors at the extremes of the training redshift
ranges we adopted the following procedure. 
For the nearby sample we trained the network using objects with spectroscopic redshift
in the range $\left[ 0.0,0.27 \right]$ and then considered the results to be reliable
in the range $\left[ 0.01,0.25 \right]$.
In the distant sample, instead, we trained the network over the range $\left[ 0.23,0.50 \right]$ 
and then considered the results to be reliable in the range $\left[ 0.25,0.48 \right]$. 

\begin{figure}
\centering
\includegraphics[width=6.5cm]{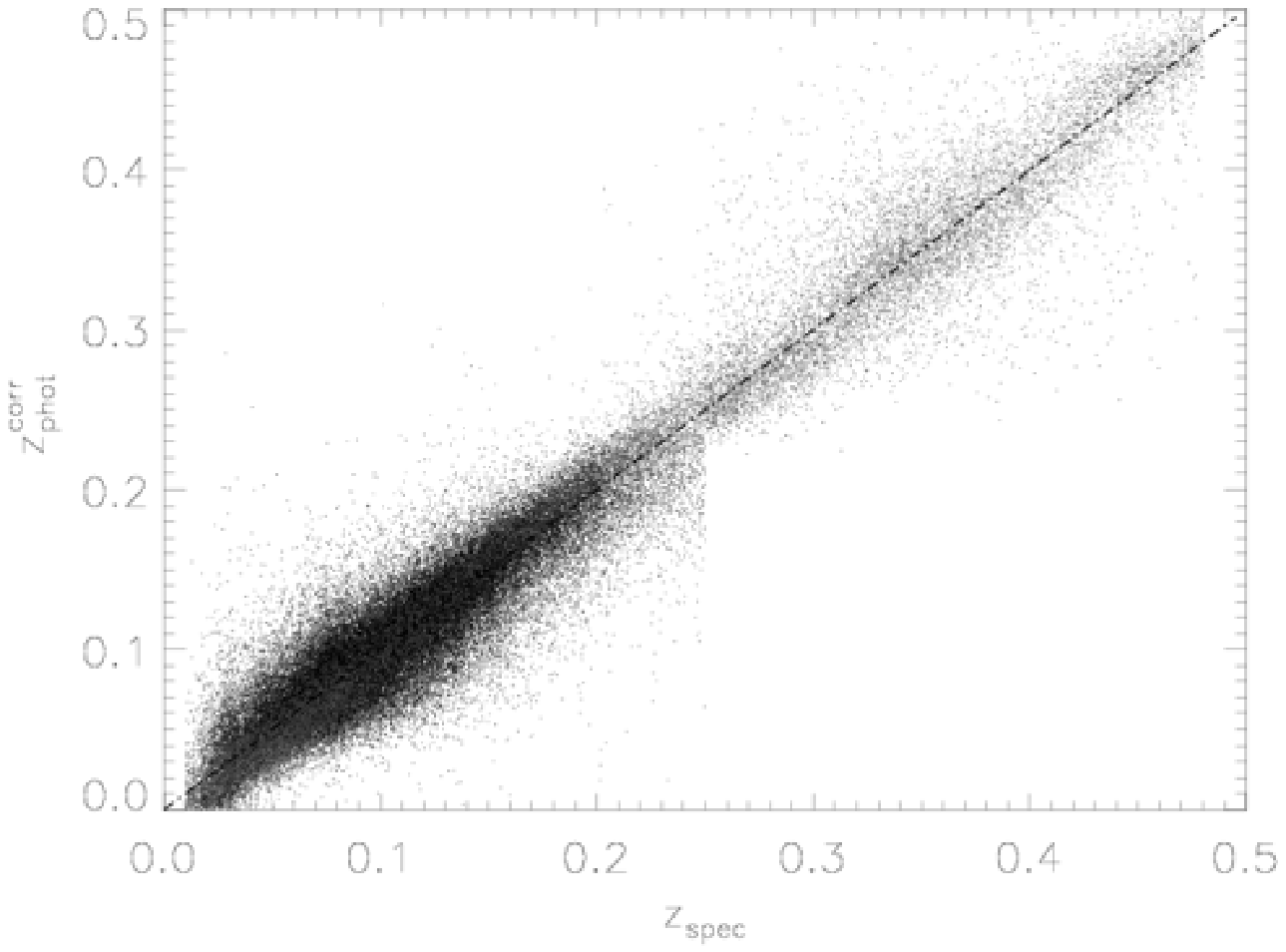}
\includegraphics[width=6.5cm]{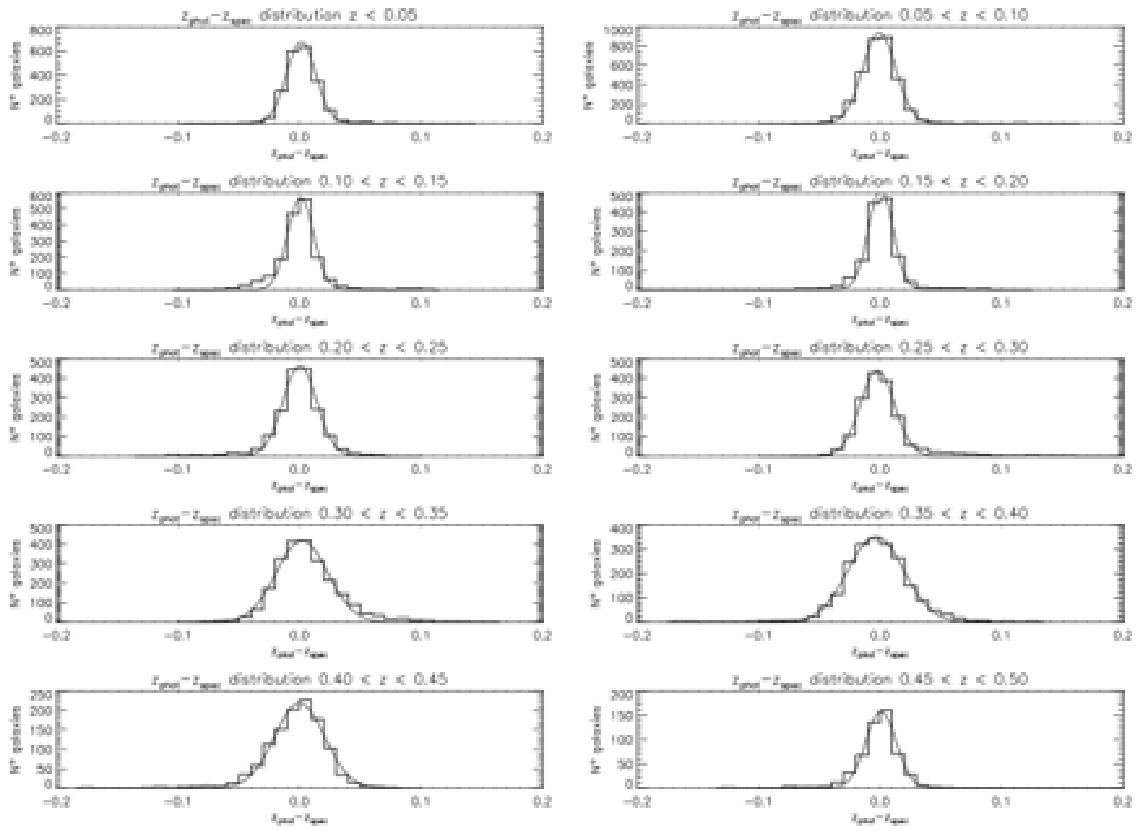}
\caption{Left: distribution of spectroscopic versus photometric redshifts in the test set. 
Lighter grey points mark LRG galaxies. Notice the larger scatter of non-LRGs in the distant 
($z>0.25$) sample. Right: distribution of the errors computed over redshift bins 
for the distant sample.}
\label{fig:redshifts}
\end{figure}

The error for our redshift estimates has been evaluated on the test set by measuring the 
dispersion of points in the $z_{spec}$ vs $ z_{phot}$ plane, i. e. the 
variance of the $z_{spec}-z_{phot}$ variable, after performing an interpolative 
correction in order to correct for systematics. 
The results can be summarized as follows:
\begin{itemize}
\item For the MG sample, in the best experiment, the robust variance turned out
to be $\sigma_3 = 0.0208$ over the whole redshift range and $0.0197$ and
$0.0245$ for the nearby and distant objects, respectively.
\item For LRG sample we obtained $\sigma_3 \simeq 0.0163$ over the whole range,
and $\sigma _3 \simeq 0.0154$ and $\sigma _3 \simeq 0.0189$ for the nearby and distant samples,
respectively.
\end{itemize}
\section{The clustering}\label{clustering}
The implementation of a reliable classification scheme based on 
photometric information only, is an old problem in astronomy 
\cite{RC2}.  
In few words, it consists in partitioning the observed parameter space
in clusters of objects sharing some underlying common physical property.
Obviously, since there is no a priori reason to assume that photometric classification
must reflect strictly any morphological classification (actually the degeneracy observed
in colours strongly argues against it), any effective classification 
method must be unsupervised, id est it must partition the photometric
parameter space using only the statistical properties of the data themselves.
We therefore implemented a hierarchical approach 
which starts from a preliminary clustering performed using as unsupervised 
clustering algorithm the so called "Probabilistic Principal Surfaces"  or PPS, and 
then makes use of the Negative Entropy concept and of a dendrogram structure 
to agglomerate the clusters found in the first phase. 
\subsection{PPS - Principal Probabilistic Surfaces}
Probabilistic Principal Surfaces (PPS) \cite{chang_2000,staiano_2003} are 
a nonlinear extension of principal components, in that each node on the PPS 
is the average of all data points that projects near/onto it. 
PPS define a non-linear, parametric mapping ${\mathbf y}({\mathbf x};
{\mathbf W})$ from a $Q$-dimensional latent space $({\mathbf x} \in
{\mathbf R}^Q )$ to a $D$-dimensional data space $({\mathbf t} \in 
{\mathbf R}^D)$, where normally $Q << D$. 
The function ${\mathbf y}({\mathbf x}; {\mathbf W})$ (defined continuous 
and differentiable) maps every point in the latent space to a point
into the data space.

\begin{figure}\label{fig:PPS}
\centering
\includegraphics[width=8.0cm]{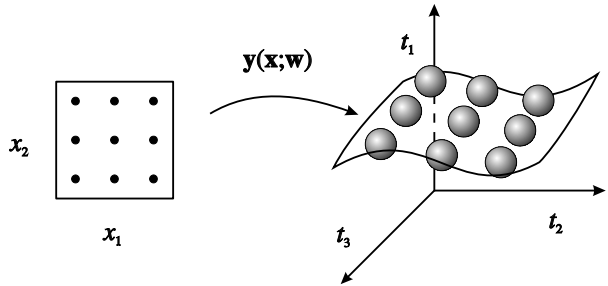}
\caption{Scheme of a latent variable model mapping a $3$D 
data space onto a $2$D latent space with latent variables uniformly 
arranged. Each latent variable is mapped, through the
nonlinear function $y$, to the center of a Gaussian distribution.}
\end{figure}
Since the latent space is $Q$-dimensional, these points will be
confined to a $Q$-dimensional manifold non-linearly embedded into
the $D$-dimensional data space. 
In our method, the points belonging to the parameter space will 
be projected on the surface of a 2-dimensional sphere. 
The visualization capabilities of the PPS can prove very useful 
in several aspects of the data interpretation phase such as, for
instance, the localization of data points lying far away from the
more dense areas (outlayers), or of those lying in the overlapping
regions between clusters, or to identify data points for which a
specific latent variable is responsible.
A visualization of the data clustering is achieved by PPS algorithm through 
the projection of objects from the multi-parametric $D$-dimesional 
space to a two dimensional surface.
These clusters, representing groups of objects corresponding to the 
same latent variable, are then decimated using  NEC: 
a hyerarchical non supervised agglomeration algorithm.

\begin{figure}\label{fig:PPS2}
\centering
\includegraphics[width=7.0cm]{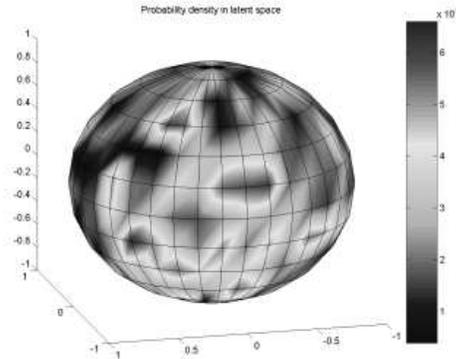}
\caption{Probability density function of the SDSS data 
plotted on the latent sphere
surface.}
\end{figure}

\subsection{NEC - Negentropy Clustering}

Most unsupervised methods require the number of clusters to be
provided \emph{a priori}. This circumstance represents a serious
problem when exploring large and complex data sets where the number of
clusters can be very high or, in any case, unpredictable. 
A simple treshold criterium is not satisfactory in most astronomical
applications due to the high degeneracy and the noisiness of the
data which lead to the erroneous agglomeration of data.

We introduce the Fisher's linear discriminant which is a
classification method that first projects high-dimensional data
onto a line, and then performs a classification in the projected
one-dimensional space \cite{bishop_1995}. 
The projection is performed in order to maximize the distance between the 
means of the two classes while minimizing the variance within each
class. 
On the other hand, we define the differential entropy H of
a random vector ${\mathbf y} = (y_1, \ldots, y_n)^T$ with density
$f({\mathbf y})$ as $ H({\mathbf y}) = \int f({\mathbf y}) \log f({\mathbf
y}) d{\mathbf y}$ so that negentropy $J$ can be defined as $
J(\mathbf{y}) = J({\mathbf y}_{Gauss}) - H(\mathbf{y})$, where
$\mathbf{y}_{Gauss}$ is a Gaussian random vector of the same
covariance matrix as $\mathbf{y}$. 
Negentropy can be interpreted as a measure of non-Gaussianity and, 
since it is invariant for invertible linear transformations, it is 
obvious that finding an invertible transformation that minimizes 
the mutual information is roughly equivalent to finding directions 
in which the Negentropy is maximized. 

Our implementation of the method makes use of an approximation to 
Negentropy which provides a good compromise between the properties 
of the two classic non-Gaussianity measures given by kurtosis and 
skeweness.
Negentropy can be used to agglomerate the clusters (regions) found 
by the PPS and the only {\it a priori} information is a dissimilarity 
threshold $T$. 
Let us suppose to have $c$ multi-dimensional regions $X_i$ with 
$i=1,\ldots,c$ that have been defined by the PPS. 
The NEC algorithm, in practice, measures whether two clusters can 
or cannot be modeled by one single Gaussian or, in other words, if the 
two regions can be considered to be just aligned or rather as a part of a 
larger data set.
By increasing the value of the dissimilarity treshold, the number of 
clusters decreases and robust partitions produce plateaus in the ''treshold 
vs number of clusters'' diagram
\begin{figure}\label{nec1}
\centering
\includegraphics[width=7cm]{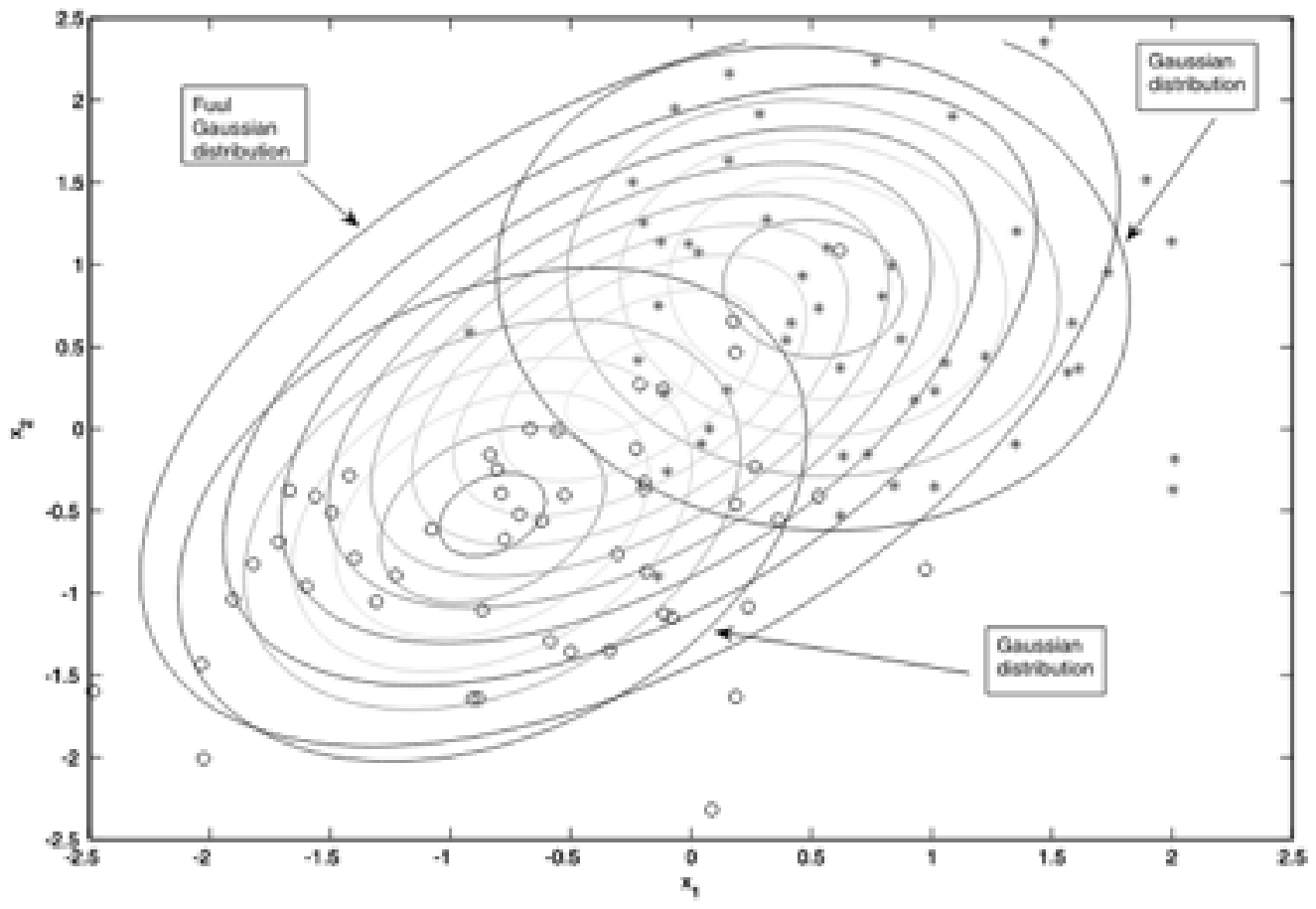}\includegraphics[width=7cm]{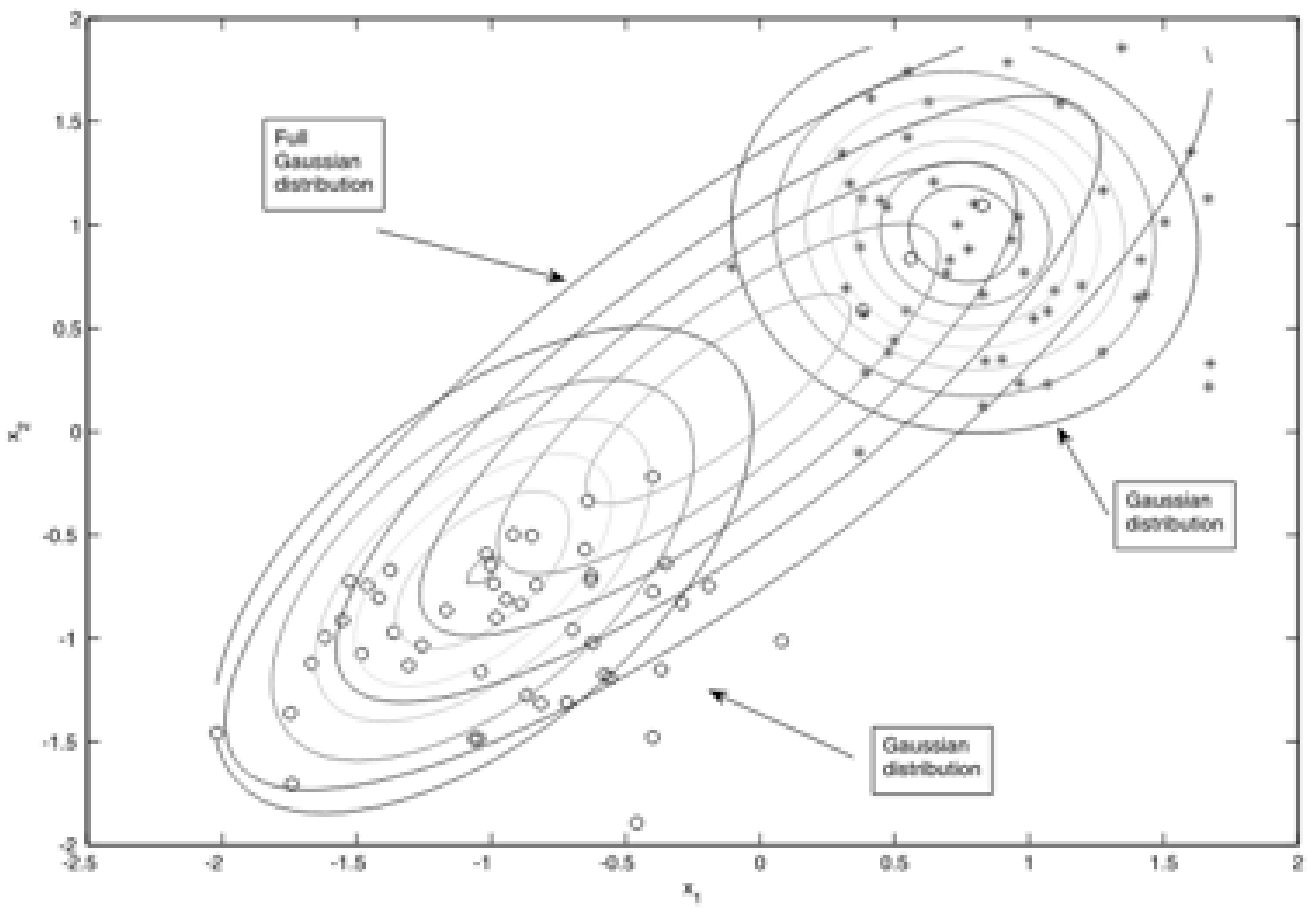}
\caption{Exemplification of how the NEC algorithm works on two 
Gaussian distribution examples. Left: Negentropy =
$2.6261$ using the $G_1$ function with $a_1 = 0.1$; Right:
Negentropy = $0.005$ using the $G_1$ function with $a_1 = 0.1$}
\end{figure}

\subsection{Preliminary results on clustering}
We first applied the PPS algorithm to the sample of 
spectroscopically selected SDSS DR-4 objects using as parameters 
for the clustering the 4 colours obtained from model magnitudes 
(u-g,g-r,r-i,i-z) of SDSS archive.
We fixed the number of latent variables and latent bases of the
PPS to 614 and 51 respectively, so obtaining at the end of this
step 614 clusters, each formed by objects which only respond to a
certain latent variables. We choose a large number of latent
variables in order to obtain an accurate separation of objects and
to avoid that any group of distinct but near points in the
parameter space could be projected onto the same cluster by chance.
The clusters found by the PPS are graphically represented
by groups of points with the same colour (a different colour for
each cluster) on the surface of a 2-d sphere embedded in the
3-dimensional latent space. 
These first order clusters were then fed to the NEC algorithm
which determined the final number of clusters. 
The plateau analysis of the agglomerative process and the inspection of the 
dendrogram allowed to set the treshold to a value corresponding to 31 clusters. 
We present in table (\ref{tabella}) the ten most populous clusters 
together with the distribution of the \emph{specClass} index within each 
cluster. The additional 21 clusters represent less than 10 $\%$ 
of the objects and are still under investigation.
It needs to be emphasized that the clustering makes use of the photometric 
data only and the spectroscopic information is used only to validate them.
As it can be seen, galaxies (SP2) clearly dominate
clusters 1, 2 and 6. 
Whether this separation reflects or not some deeper differences among the 
three groups (such as, for instance, different morphologies), cannot be 
assessed on the grounds of presently available data. 
AGNs (SP3) dominate clusters 5 and 9 even though some contamination from
galaxies  exists. 

\begin{table}
\begin{center}
\label{tabella}
\begin{tabular}{@{} c| c c c c c c c }
\hline
 Cl. n    & SP0  & SP1& SP2& SP3& SP4& SP6 \\
\hline
1 &69 &145 &9362 &48 &0 &12\\
2 &25 &133 &13370 &10 &0 &12\\
3 &149 &132 &63 &64 &0 &5\\
4 &44 &3396 &1530 &189 &67 &1\\
5 &202 &85 &447 &2428 &6 &10\\
6 &26 &125 &13728 &12 &0 &12\\
7 &0 &0 &0 &0 &0 &484\\
8 &1 &1 &1 &0 &0 &329\\
9 &541 &1507 &127 &4750 &18 &1\\
10 &89 &474 &2117 &19 &4 &529\\
\hline
\end{tabular}
\end{center}
\caption{The table shows the distribution of objects 
in the most significant clusters found by our procedure. 
The columns correspond to different values of the 
\emph{specClass} index.}
\end{table}

Late type stars (SP6) populate mainly clusters 7 and 8 and are strong contaminants 
of cluster 10 which also is dominated by  galaxies. 

\section{Conclusions}\label{conclusions}
We have produced a catalogue of redshifts for all galaxies in 
the SDSS database matching the following selection criteria and
falling in the redshift range $0.005 - 0.5$. 
These redshifts have been used to produce a first 
3-D map of the nearby universe which is currently under study
to identify structures and to define accurate selection functions. 
Much work has instead still to be done
in obtaining a reliable partition in hogeneous types. This is partly due 
to the lack of a reliable data sets for labeling and partly to a 
still poor understanding of how to deal in an effective way with 
upper limits and missing data.\\

\noindent {\it Acknowledgements:} this work was sponsored by the
Italian MIUR through a PRIN grant and by the European Union through the VO-Tech 
project. The authors wish to thank G. Miele, G. Raiconi and N. Walton for 
useful discussions.


\begin{thebibliography}{9}
\bibitem{votech} WWW site: {\it http://eurovotech.org}.
%\bibitem{astronomy} Blake C., Bridle S., 2005, MNRAS, 363-4, 1329-1348.
\bibitem{longo_2005} Staiano A., De Vico L., Ciaramella A., Donalek C., Longo G., Raiconi G., 
Tagliaferri R., Amato R., Del Mondo C., Mangano G. \& Miele G., 2005, IEEE, in print.
\bibitem{SDSS5} Adelman-McCarthy J., et al., 2007, ApJS, submitted.
\bibitem{dabrusco_2006} D'Abrusco R., Staiano A., Longo G., Brescia M., De Filippis E., 
Paolillo \& Tagliaferri R. 2006, ApJ, submitted.
\bibitem{tagliaferri_2003} Tagliaferri R., Longo G., Andreon S., Capozziello S.,
Donalek C., \& Giordano ,G., 2003, Lectures Notes in Computer Science, 2859, 226.
\bibitem{bishop_1995} Bishop C. M., 1995, {\it Neural Networks for Pattern Recognition}, 
Oxford University Press.
\bibitem{duda} Duda R.O., Hart P.E., Stork D.G., 2001, {\it Pattern
Classification}, John Wiley \& Sons Inc., Second Edition.
\bibitem{RC2} de Vaucouleurs G, de Vaucouleurs A., Corwin H.C. jr, 1976, {\it Second Reference 
Catalogue of Bright Galaxies}, University of Texas Press.
\bibitem{chang_2000} Chang K., 2000, {\it Nonlinear Dimensionality Reduction Using
Probabilistic Principal Surfaces}, PhD Thesis, The University of Texas at Austin (USA).
\bibitem{staiano_2003} Staiano A. 2003, {\it Unsupervised Neural Networks for the
Extraction of Scientific Information from Astronomical Data}, PhD thesis, University of Salerno (Italy).

%\bibitem{Mazu84} O.V. Mazurin and E.A. Porai-Koshits (eds.),
%                 Phase Separation in Glass, North-Holland, Amsterdam, 1984.
%\bibitem{Dimi75} Y. Dimitriev and E. Kashchieva, 
%                 J. Mater. Sci. 10 (1975) 1419.
%\bibitem{Eato75} D.L. Eaton, Porous Glass Support Material,
%                 US Patent No. 3 904 422 (1975).
\end{thebibliography}
\end{document}